\documentclass[reprint,twocolumn,showpacs,showkeys]{revtex4-1}

\usepackage{latexsym,graphicx,graphics}
\usepackage{amsmath,epsfig}

\newcommand{\abs}[1]{\left| #1 \right|}

\newcommand{\bb}[1]{ \mbox{\boldmath$ #1$}}

\newcommand{\rv}{\bb r}

\newcommand{\Eq}[1]{Eq.~(\ref{#1})}

\newcommand{\unit}[1]{\bb{\hat{#1}}}

\begin{document}

\title{One-way optical waveguides for perfectly matched non-reciprocal nano-antennas with dynamic beam tuning}
\author{Y. Hadad}
\surname{Hadad}
\email{hadady@eng.tau.ac.il}
\author{Ben Z. Steinberg}
\surname{Steinberg}
\email{steinber@eng.tau.ac.il}\thanks{corresponding author}
\affiliation{School of Electrical Engineering, Tel Aviv University, Ramat-Aviv, Tel-Aviv 69978  Israel}
\date{October 2010}

 \begin{abstract}
 An optical signal that propagates along a conventional waveguide is back reflected at waveguide terminations. If the waveguide permits only one-way propagation, this back reflection cannot take place; the optical signal undergoes a nearly 100\% conversion to radiation modes. Hence, a terminated one-way waveguide may serve as a matched nano-antenna with a natural guiding and feeding mechanism. We examine the termination effect in a recently suggested nano-scale one-way plasmonic waveguide, and its properties as a matched nano antenna are shown. Since one-way structures are inherently non-reciprocal, our antenna possesses different transmitting and receiving patterns, and a dynamic beam-forming functionality. These unique properties offer new possibilities for dynamic directional selectivity that differ in transmit and receive operations. Finally, the new concept offers nano-antennas as yet another important field of application for non-reciprocal nano-photonics, accelerating research of both the former and the latter fields of activity.

 \end{abstract}


\maketitle

Antenna theory in the radio-frequency (RF) regime is a well established discipline, whose overwhelming practical success can be attributed mainly to its conceptually modular nature \cite{AntennaBook}. A RF signal is generated at a given space-point by an oscillator. It is then transferred by a waveguide to an antenna whose \emph{remote} location is determined by system considerations. The waveguide is connected to the antenna port via a feeding mechanism or matching circuit whose role is to eliminate back reflections that arise at the antenna port if the waveguide and the antenna impedances mismatch. Antenna characterization concepts such as directivity and gain are then used to determine the radiated signal intensity, spatial properties and beam quality. Since in most RF applications the antenna and its surroundings are reciprocal, reciprocity theorem establishes symmetry between the system properties at transmit (Tx) and receive (Rx) modes.

An effort is devoted recently to the emerging field of nano-antennas designed to operate at optical or IR wavelengths \cite{Hanson_IntegEq_IEEEAP_2006,Novotny_Scaling_PRL_2007,AluEngheta_nphoton_2008,AluEngheta_PRL_2008,
AluEngheta_PRB_2008,Cozier_nphoton_2009,YagiUdaNaturePhotonics,Novotny_nphoton_2011,
Ramaccia_OpLett_2011,Alu_LeakyWaveChainAntenna_PRB_2010,Lomakin2,Lomakin3}. It is motivated by potential applications in diverse fields such as energy harvesting and concentration, local lighting or heating, wireless optical links, sensing, detection, and more. Despite the advancement in nano-scale production technologies, a mere down-scale of conventional antennas is not feasible due to fabrication limitations and the profound change of metal properties when the frequency is increased to the optical regime. Hence, the modular nature of RF antennas discussed above cannot be directly applied. New mathematical modeling and scaling tools of antenna elements that take into account material dispersion were developed \cite{Hanson_IntegEq_IEEEAP_2006,Novotny_Scaling_PRL_2007} and applied for various classical as well as novel antenna geometries \cite{Novotny_nphoton_2011}. Other efforts are based on ad-hoc tuning of classical geometries (e.g.~nano-dipoles, Yagi-Uda antennas, nano-dimers) achieved by various antenna loading strategies, or geometrical tuning \cite{AluEngheta_nphoton_2008,AluEngheta_PRL_2008,AluEngheta_PRB_2008,Cozier_nphoton_2009,YagiUdaNaturePhotonics}. Leaky wave nano-antennas based on particle chains were considered too, both in regular \cite{Alu_LeakyWaveChainAntenna_PRB_2010} and chiral \cite{Lomakin2,Lomakin3} geometries. In all these previous studies the antenna resonance and radiation pattern were investigated. However, the issues of source to nano-antenna energy transport, e.g.~guiding and matching, remain largely unexplored; the antenna radiates right from the source location (an exception is the horn antenna fed by a parallel plates waveguide \cite{Ramaccia_OpLett_2011}, but its size is of several microns.)

\begin{figure}[htbp]
\vspace*{-0.in}
    \hspace*{-0.1in}
        \includegraphics[width=7.5cm]{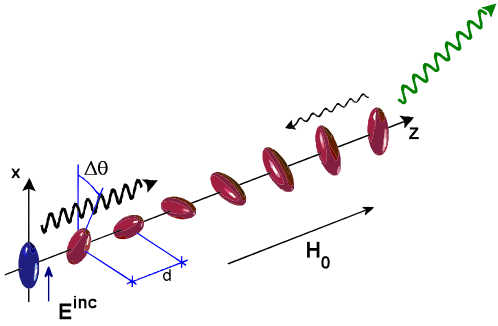}\vspace*{-0.1in}
    \caption{A matched nano-antenna based on a terminated one-way waveguide. The port (blue ellipsoid) is excited by a local field $\bb{E}^{\mbox{\tiny inc}}$, generating a guided mode that propagates in the (allowed) $+z$ direction. Guided mode back reflections at the remote end cannot take place, hence it is converted to radiation. This end is a matched local antenna.}
    \label{fig1}
\end{figure}

  Here we suggest a \emph{new concept} for source-waveguide-nannoantenna assembly that, although not modular, encapsulates all the three important components together in a perfectly matched and natural way. Thus it may offer a \emph{new generation} of nano-antennas free from the traditional difficulties articulated above. In our scheme, shown in Fig.~\ref{fig1},
a terminated one-way plasmonic waveguide serves as the key component. Consider a guided mode that propagates, say in the $+z$ direction, in a conventional waveguide of finite length. Usually when it hits the waveguide end most of its energy is reflected back into the waveguide. However, if it is a one-way waveguide (i.e.~it supports propagation only in the $+z$ direction) these back reflections cannot take place so the guided mode is converted to radiation (free-space) modes with nearly 100\% efficiency. This conversion to radiation mode takes place only along a limited region near the end, that plays the role of a local antenna. Furthermore, since one-way structures are inherently non-reciprocal, the antenna possesses different transmitting and receiving patterns, and a dynamic beam-forming functionality.

We emphasized that this work is not about new one-way waveguides per-se. Rather, we use a recently suggested nano-scale one-way waveguide \cite{Hadad_Steinberg_PRL} in order to demonstrate the efficacy of the new concept discussed above. This one-way plasmonic waveguide is based on the interplay between two physical phenomena: the non-reciprocal optical Faraday (or cyclotron) rotation, and structural chirality. The former is obtained by exposing a sub-diffraction chain (SDC) of plasmonic particles to a longitudinal DC magnetic field $\bb{H}_0$. The latter is obtained by constructing the SDC with ellipsoidal particles arranged as a spiral. The two-type rotations interplay enhances the non-reciprocity and leads to one-way guiding within a finite frequency band. The structure is shown in Fig.~\ref{fig1}. Similar structures were fabricated, e.~g.~in\cite{Scherer} for different applications. The system consists of $N$ prolate ellipsoids, the $m$-th particle
centered at $\rv_m=(0,0,md)$, and rotated by $m\Delta\theta$ about the $\unit{z}$ axis. The entire structure is subject to an external DC magnetic field $\bb{H}_0=\unit{z}H_0$. Under the discrete dipole approximation (DDA), the chain response is governed by the $3N\times 3N$ matrix equation \cite{Hadad_Steinberg_PRL}
\begin{equation}
\bb{p}_m -\epsilon_0^{-1}\bb{\alpha}_m\sum_{n\ne m}^N{\bf A}[(m-n)d]\bb{p}_n
           = \bb{\alpha}_m\bb{E}^{\mbox{\tiny inc}}(\rv_m)\label{eq1}
\end{equation}
where $\bb{\alpha}_m$ and $\bb{p}_m$ are the $m$-th particle polarizability and dipole moment response, respectively. $\bb{\alpha}_m$ is given by
\begin{equation}
\bb{\alpha}_m={\bf T}_{-m}\bb{\alpha}{\bf T}_m, \label{eq2}
\end{equation}
where ${\bf T}_m$ is a transverse rotation matrix, whose non-zero entries are $t_{11}=t_{22}=\cos m\Delta\theta$, $t_{33}=1$. $t_{12}=-t_{21}=\sin m\Delta\theta$,
$\bb{\alpha}$ is the matrix polarizability of the reference prolate ellipsoid (say, the blue ellipsoid at the origin in Fig.~\ref{fig1}), subject to a longitudinal magnetic field $\bb{H}_0=\unit{z}H_0$ - see \cite{Hadad_Steinberg_PRL} for details. The matrix ${\bf A}(z-z')$ represents the electric field at $(0,0,z)$ due to a short dipole $\bb{p}$ at $(0,0,z')$,
\begin{eqnarray}
\bb{E}(z)&=&\epsilon_0^{-1}{\bf A}(z-z') \,\bb{p}\label{eq3}\\
{\bf A}(z)&=&\frac{e^{ik\abs{z}}}{4\pi\abs{z}}\,\left[k^2{\bf A}_1 + \left(\frac{1}{z^2}-\frac{ik}{\abs{z}}\right){\bf A}_2\right]
\label{eq4}
\end{eqnarray}
here ${\bf A}_1=\mbox{diag}(1,1,0),\,{\bf A}_2=\mbox{diag}(-1,-1,2)$. Once \Eq{eq1} is solved for the dipole moments $\bb{p}_m$, one can obtain the field at any point in space by a superposition of short dipole radiation fields.

 When the system is viewed as an antenna, some entities need to be defined. We refer to the blue ellipsoid particle at the origin as the \emph{system port}. In Tx mode, the system is excited by a local source whose electric field $\bb{E}^{\mbox{\tiny inc}}(\rv_m)=\unit{x}E^{\mbox{\tiny inc}}\delta_{m,1}$ is induced on the port. The input power transmitted by the source to the entire structure, including the power that radiates to the far field region, is given by
\begin{equation}
P_{\mbox{\tiny in}}=\mbox{Re}\left\{ \bb{E}^{\mbox{\tiny inc}}\cdot i\omega \bb{p}_1^*\right\}\label{eq5}
\end{equation}
where $\bb{p}_1$ is the dipole moment excited at the port-ellipsoid, and $^*$ denotes complex conjugate. A measure of the feeding quality, i.e. the matching between the port and the antenna radiating elements (in the other end of chain), is $\Gamma_{11}$ - the return loss. It is defined as the ratio between the chain modes that propagate in the $+\unit{z}$ and $-\unit{z}$ directions. It can be obtained by applying a discrete Fourier transform (DFT) on the chain dipole moments $\bb{p}_n^r={\bf T}_n\bb{p}_n$, and computing the ratio between the spectral components that correspond to left and right propagations (${\bf T}_n$ is present to eliminate spurious side-lobes at $\pm\Delta\theta$ due to the structure rotation).
Finally, the antenna \emph{gain} in the direction $\unit{r}$, $\bb{G}_{\mbox{\tiny T}}(\unit{r})$, is defined as the ratio
\begin{equation}
\bb{G}_{\mbox{\tiny T}}(\unit{r})=\bb{S}^{\mbox{\tiny FF}}(\unit{r})/\max_{\theta,\phi}\bb{S}_{\mbox{\tiny sd}}^{\mbox{\tiny FF}}(\unit{r})\label{eq6}
\end{equation}
where $\bb{S}^{\mbox{\tiny FF}}(\unit{r})$ and $\bb{S}_{\mbox{\tiny sd}}^{\mbox{\tiny FF}}(\unit{r})$ are the far-field Poynting vector radiated by our antenna and by an antenna that consists of a single particle identical to the port (short dipole) respectively, \emph{both with the same incident field} $\bb{E}^{\mbox{\tiny inc}}$ - i.e.~with the same input port voltage. Hence, the gain includes also a measure of the matching between an exciting oscillator and our antenna assembly.

 In Rx mode, the system is excited by an incident plane wave that illuminates the entire chain, and the dipole moment excited at the system port is referred to as the ``received signal''. The antenna gain is then given by
 \begin{equation}
 G_{\mbox{\tiny R}}=\abs{\bb{p}_1}^2/\abs{\bb{p}_{\mbox{\tiny sd}}}^2 \label{eq7}
 \end{equation}
where $\bb{p}_1$ is the dipole moment response at the antenna port, and $\bb{p}_{\mbox{\tiny sd}}$ is the dipole moment response of a single identical particle under the same incident field.

\begin{figure}[htbp]
\vspace*{-0.in}
    \hspace*{-0.2in}
        \includegraphics[width=7.cm]{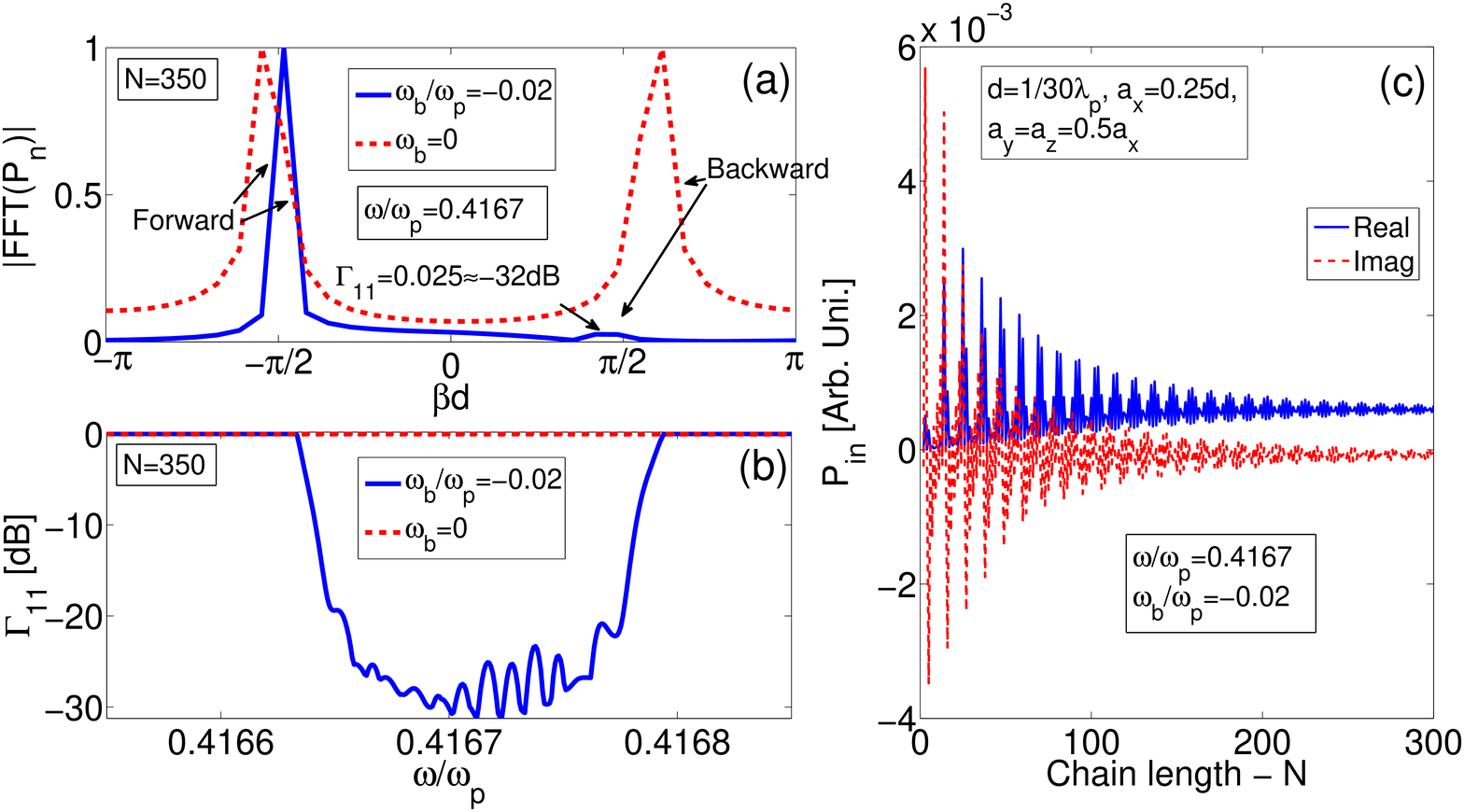}\vspace*{-0.1in}
    \caption{Antenna matching in Tx mode. (a) An FFT of the chain response for a frequency inside the one-way band. (b) The return loss near the input port, vs. frequency. (c) Transmitted power vs. chain length.}
    \label{fig2}
\end{figure}

We turn now to demonstrate the antenna operation of this structure, using the entities defined above. Specifically, we concentrate in the one-way lower band \cite{Hadad_Steinberg_PRL}, with the following parameters. The ellipsoids semi-axis are $a_x=0.25d, a_y=a_z=0.5a_x$ and $d=\lambda_p/30$, where $\lambda_p$ is the $\omega_p$ wavelength. The rotation step is $\Delta\theta=70^\circ$. The magnetic field strength and sign ($\pm\unit{z}$ direction) correspond to the cyclotron frequency $-q_eB_0/m_e\equiv\omega_b=\pm 0.02\omega_p$, respectively. With these parameters, one-way guiding in the $\mp\unit{z}$ direction  is created at a finite bandwidth around the frequency $\omega=0.4167\omega_p$. The group and phase velocities are in opposite sign. We have excited a chain with 350 particles in the Tx mode, and solved \Eq{eq1}. Figure \ref{fig2}(a) shows the FFT of $\bb{p}^r_n$.
When $\omega_b=0\Rightarrow H_0=0$ (one way property is turned off), the spectrum is symmetrical around the origin, indicating that guided mode to guided mode back reflection of nearly 100\% takes place. The peaks location correspond exactly to the guided mode (real) roots of the chain determinant studied in \cite{Hadad_Steinberg_PRL}, when no magnetic field is present. When $\omega_b=-0.02\omega_p$ (the one way property is turned on), the peaks shift to non-symmetrical locations. Now, the location of the left (right) peak correspond exactly to the real (real part of the complex) root of the one-way chain determinant \cite{Hadad_Steinberg_PRL}, representing $+\unit{z}$ ($-\unit{z}$) propagating guided (radiation) mode. With the presence of magnetic field the back reflection peak is reduced considerably; this is because the back reflected mode is a radiation mode that looses its energy very fast as it radiates to the free space. The return loss, given as the ratio between the backward and forward peaks, is shown for a band of frequencies in Fig.~\ref{fig2}(b). The antenna is well matched over the entire one-way regime. Another measure of the source-to-antenna matching is the power transmitted to the system $\bb{E}^{\mbox{\tiny inc}}\cdot i\omega \bb{p}_1^*$ as a function of the chain number of particles $N$ (chain length), shown in Fig.~\ref{fig2}(c). For small $N$, the chain is shorter than the radiation mode decay length, hence the two ends may resonate and produce the typical oscillating behavior of the transmitted power. For $N>100$ the curves level off, indicating a net supply of real power and a near-zero supply of reactive power.

\begin{figure}[htbp]
\vspace*{-0.05in}
    \hspace*{-0.1in}
        \includegraphics[width=7.5cm]{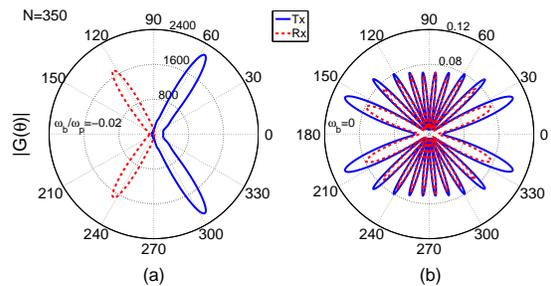}\vspace*{-0.1in}
    \caption{Antenna Tx and Rx gain in the $xz$ plane ($\phi=0$). (a) One way is on. (b) One way is off. The gain in $yz$ plane is the same. Here $N=350$.}
    \label{fig3}
\end{figure}

\begin{figure}[htbp]
    \centering
    \hspace*{-0.1in}
        \includegraphics[width=7.6cm]{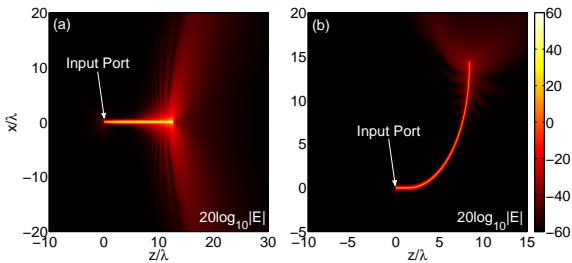}\vspace*{-0.05in}
    \caption{Tx antenna $E$-field in $xz$ plane, using a straight (a) or curved (b) chain. Radiation is emitted essentially from the chain end. Curved geometries may be used to overcome chip-design constraints.}
    \label{fig4}
\end{figure}

\begin{figure}[htbp]
\vspace*{-0.1in}
    \hspace*{-0.2in}
        \includegraphics[width=7cm]{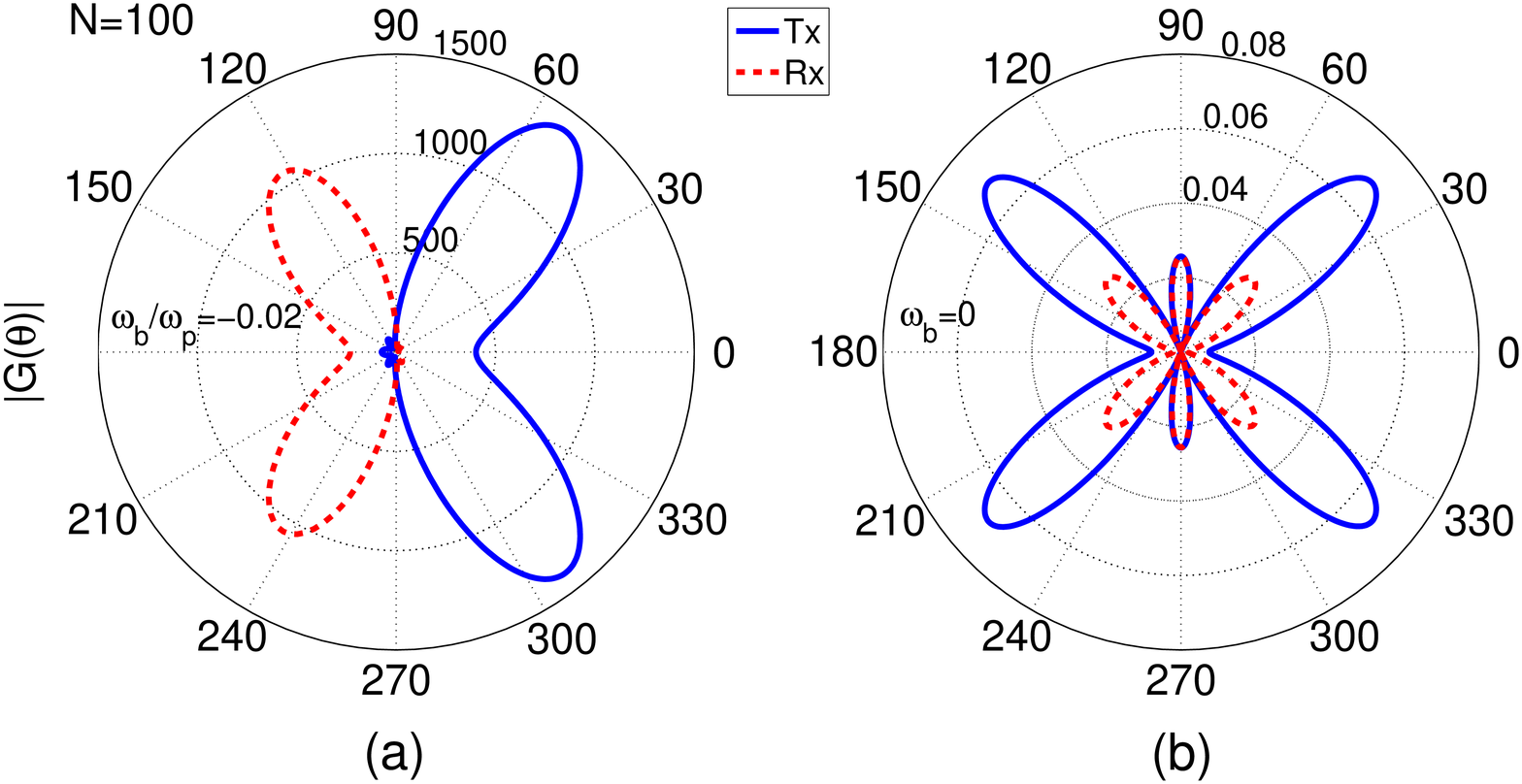}\vspace*{-0.1in}
    \caption{The same as Fig.~\ref{fig3}, for $N=100$.}
    \label{fig5}
\end{figure}

The antenna gain in both Tx and Rx modes are shown in Fig.~\ref{fig3}(a) with the one-way property, and in Fig.~\ref{fig3}(b) without it (turning off the magnetic field), for the frequency $\omega=0.4167\omega_p$.
Clearly, in the former the gain and beam quality are very high, and in the latter they reduce considerably. Additional interesting observations can be made. Since the one-way property is inherently non-reciprocal, the Tx and Rx gain curves in Fig.~\ref{fig3}(a) point at different directions; the antenna transmits at one direction, and receives from another. These directions correspond precisely to the real part of the complex roots of the one-way chain determinant \cite{Hadad_Steinberg_PRL}. When one-way is off, the system is reciprocal hence Rx and Tx gain possess the same directional properties, as seen in Fig.~\ref{fig3}(b). With the present parameters, chain length is $4.8\lambda$; the observed multi-lobe low gain is due to the chain length and the fact that for conventional (two-way) chain the reflection magnitude at the chain terminations is nearly a unity \cite{HadadSteinberg_PRB}. This is typical to long wires radiation. The fact that with the one-way property the beam is collimated and possesses high gain, is due to the near perfect matching and due to the coherent radiation mode that radiates essentially from the chain end. This is seen in E-field plot of Fig.~\ref{fig4}. Essentially, a section of $\simeq 100$ particles ($\simeq 1.3\lambda$) at the chain end radiates to the free space, for both straight or curved geometries. As a direct result, we have observed that the gain does not change when the chain length is increased.
However, when the length is dramatically decreased, gain reduces. Figure \ref{fig5} is the same as Fig.~\ref{fig3}, but for a chain of 100 particles only.

\begin{figure}[htbp]
\vspace*{-0.in}
        \includegraphics[width=7cm]{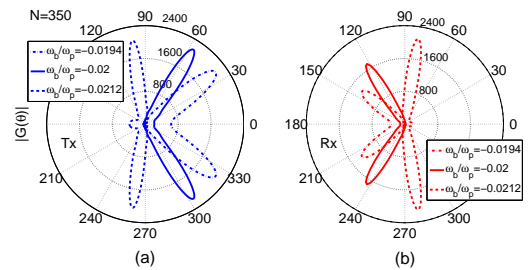}\vspace*{-0.1in}
    \caption{Dynamic control of the optical beam by the bias magnetic field. (a) Tx gain. (b) Rx gain. 18\% variation of $H$ provides $60^\circ$ scanning range.}
    \label{fig6}
\end{figure}

Finally it is important to note that the bias magnetic field can be varied over quite a large range without deteriorating the one-way property. However, this variation induces a considerable change of the radiation mode complex wave-number (chain determinant roots \cite{Hadad_Steinberg_PRL}). Hence, our system possesses a \emph{dynamic beam tuning functionality}. A change of few percents of $H_0$ causes a considerable variation of Tx and Rx directions. This is demonstrated in Fig.~\ref{fig6}.

We conclude our discussion by commenting that the concepts presented here may be applied to any one-way guiding scheme, e.g. \cite{YUFAN}, leading to potentially perfect matching of nano-antennas to feeding and guiding systems.

\end{document}